\documentclass[12pt]{article}
\newcommand{\bos}{^{\bf B}}
\newcommand{\ferm}{^{\bf F}}
\newcommand{\rep}{{\cal G}}
\newcommand{\eqn}[1]{(\ref{#1})}

\newcommand{\ft}[2]{{\textstyle\frac{#1}{#2}}}

\renewcommand{\u}[1]{{\bar{#1}}}
\newcommand{\ba}{\left(\begin{array}}
\newcommand{\ea}{\end{array}\right)}

\newcommand{\M}{{\cal M}}
\newcommand{\R}{{\cal R}}
\newcommand{\T}{{\cal T}}

\newcommand{\E}{{\Upsilon}}
\renewcommand{\L}{{\cal L}}

\begin{document}
\begin{titlepage}
\begin{flushright}
SU-ITP-98/61\\
KUL-TF-98/56\\
{\tt hep-th/9812087}\\
December 10 1998
\end{flushright}
\vfill
\begin{center}
{\Large\bf Superisometries of the adS$\,\times\,$S Superspace\\[20pt]}
\vskip 0.3cm
{\large {\sl }}
\vskip 13.mm
{\bf  Piet Claus$^+$  and ~Renata Kallosh$^{*}$}\\
\vskip.5cm
{\small
$^+$
Instituut voor theoretische fysica, \\
Katholieke Universiteit Leuven, B-3001 Leuven, Belgium\\
{\it Piet.Claus@fys.kuleuven.ac.be}\\[8pt]
\
$^*$ Physics Department, \\
Stanford University, Stanford, CA 94305-4060, USA\\
{\it kallosh@physics.stanford.edu}
}
\end{center}
\vskip 2cm
\begin{center}
{\bf Abstract}
\end{center}
\begin{quote}
We find the superisometry of the near-horizon
superspace, forming the superconformal algebra.  We
present here the explicit form of the transformation of the bosonic and
fermionic coordinates (as well as the compensating Lorentz-type
transformation) which keeps the geometry  invariant. We comment on  
i) the BPS
condition of the branes in adS background and  ii) significant  
simplification
of  superisometries of the gauge-fixed Green-Schwarz action in  
$adS_5\times
S^5$  background.
\end{quote}
\vfill
\end{titlepage}
\section{Introduction}
It has been suggested in \cite{GT} to consider the branes as interpolating
solutions between two maximally supersymmetric vacua.  The first vacuum is
related to the brane at infinity and is given by a flat space without
forms.  Flat space has the maximal amount of unbroken supersymmetry
and can
be understood as a flat superspace.  It has a description \cite{SS} in
terms of the supercoset construction $G/H$ where $G$ is the super
Poincar\'{e} group and $H$ is its Lorentz subgroup, see also \cite{West}
for details.  The second vacuum is associated with the near-horizon
geometry of the brane solutions, given by the $adS_{p+2}\times S^{d-p-2}$
metric and a form field.  It can also be understood as a superspace since
the amount of unbroken supersymmetries is maximal.  Recently a supercoset
$G/H$ construction \cite{MT, KRR, Bernard} of the near-horizon
superspace $adS_{p+2}\times S^{d-p-2}$ + form geometries was developed.
Here $G$ is the relevant (extended) superconformal group and the stability
group $H$ is the product subgroup $SO(p+1,1)\times SO(d-p-2)$.  It
was also
shown that the supercoset construction of the near-horizon superspace is
equivalent to a description in terms of the supergravity superspace at the
fixed point where all covariant geometric superfields become covariantly
constant \cite{KRaj,C}.  These vacua are exact as shown in \cite{KRaj} in
the framework of supergravity.  One can also understand the exactness
property by the fact that the relevant supercoset construction is
completely defined by the supergroup $G$.
\par
We will develop here the supercoset construction of these superspaces with
the purpose to find the superisometries of the $adS \times S$ superspace.
All equations will be kept in a form valid for the super Poincar\'{e} as
well as for the superconformal group.  This will be useful for a simple
check of the consistency of our construction since in case of the flat
superspace the superisometries are known.  To explain our goal in the
$adS\times S$ case we recall below the superisometries of the flat
superspace.
\par
The flat superspace with coordinates $Z =\{ x,\theta\}$ is defined by
the vielbein superforms
\begin{equation}
E^a \equiv dx^\mu \delta_\mu^a + \bar \theta \gamma^a d \theta \ ,  \qquad
E^\alpha \equiv d\theta^\alpha \ .
\end{equation}
The isometries of the flat superspace are defined by the invariance of the
superspace vielbeins, under the change of coordinates of the superspace
$\delta Z = -\Xi(Z)$ supplemented by a compensating Lorentz transformation
with the parameter $\Lambda^{ab}= \lambda^{\mu\nu}_M \delta^a_\mu
\delta^b_\nu$
on the vielbeins, i.e.
\begin{equation}
\Delta  E \equiv {\cal L}_{\Xi} E + \delta_{Lor}(\Lambda) E =0 \
 \ .
\end{equation}
One finds that the superisometries are given by
\begin{equation}
\Xi (Z) : \qquad - \delta\theta =  \epsilon + {1\over 4}\lambda^{\mu\nu}_M
\gamma_{\mu\nu} \theta \ , \qquad  - \delta x^\mu = a^\mu +
\lambda^{\mu\nu}_M x_\nu
+ \bar \epsilon  \gamma^\mu \theta \ .
\label{flatisom}
\end{equation}
with $x, \theta$-independent $a^\mu, \lambda^{\mu\nu}, \epsilon$,
which are the
parameters of a global Poincar\'{e} supersymmetry.  This set of symmetries
is known to be present in the classical actions of the
$GS$-superstring\footnote{In particular, the fact that the spectrum of
states of $GS$
string upon quantization forms the representations of the Poincar\'{e}
supersymmetry can be traced back to the superisometry of the
background.}, $BST$-supermembrane, $M5$-brane, all $D$-brane actions
etc.~as long as the
background is a flat superspace.  This symmetry follows from the fact that
the worldvolume actions depend on the pull-back to the worldvolume of the
vielbein forms $E$.  The equation above states that the actions of the
extended objects in the flat superspace background have manifest super
Poincar\'{e} symmetry.
\par
We would like to exhibit the analogous manifest superconformal symmetry of
the actions of the extended objects placed in the $adS\times S$ + form
background.  One of the motivation is to use these symmetries to find the
spectrum of states e.g.~of the $GS$ superstring in $adS_5\times S^5$
background \cite{GS,PES}
which will form the representations of $SU(2,2|4)$ supergroup.
For this purpose we will find the analogous transformations
$\Xi(Z), \Lambda^{ab}(Z)$ for the near-horizon superspace under which
the vielbein and the connection forms and therefore the classical actions
are invariant.
\par
The paper presents a new $G$-covariant approach to the definitions of the
superisometries of the ${G/H}$ supercoset space.  This approach is  
based on
the concept of the gauged \cite{Toinekar} superalgebra {\bf G} and is
described in Sec.~2.  It allows us to formulate the $G$-covariant Killing
equations in superspace and solve them to all orders in fermionic
coordinates of the superspace $\theta$ in terms of the values of these
Killing fields at $\theta=0$.  Our construction is independent of a
coordinate choice for the bosonic space (the superspace at $\theta=0$),
which is encoded in the choice of a coset representative of the bosonic
subgroup.  In Sec.~3 we show how to derive from the $G$-covariant Killing
vectors in superspace the standard form of superisometries, i.e.~the
transformation of the coordinates of the supercoset space and the
compensating transformation of the stability group $H$.  As a warm up we
apply our general formulae to the flat superspace in Sec.~4 before we
specialize to the particular case of interest, the $adS_{p+2}\times  
S^{d-p-2}$
superspaces. We supplement our general formulae with the relevant  
superalgebra
and the Killing fields, i.e. Killing vectors and spinors, of the  
bosonic space
in coordinates in which the metric
is a product metric.
The bosonic part of the superconformal isometries was found in  
\cite{Malda}
for the case of only radial excitations on the worldvolume and extended to
the most general bosonic case in \cite{conf}.  Also the Killing  
spinors for
$adS\times S$ geometry are known \cite{LPR}.  We add here the relevant
$\theta=0$ expressions for the $H$ transformations.
\par
The $adS\times S$ superspace has been obtained in the context of brane
solutions to supergravity.  In supergravity the Killing equations are the
transformations that leave the supergravity solution invariant.  In
particular the Killing-spinor equation is the vanishing transformation of
the fermionic fields at vanishing fermionic fields.  In supergravity the
group structure of the superspace is not manifest.  We show how to derive
the superisometries in supergravity superspace by interpreting the
curvature and torsion constraints as Maurer Cartan equations.
Particularly when the metric is not a product space, one may require this
alternative description.  An example is when the $adS\times S$ metric is
rewritten in cartesian coordinates, in which the metric is invariant with
respect to the directions along the brane and to those transverse to the
brane.  The R-symmetry of the conformal group is manifest in these
coordinates.  The worldvolume actions of various branes are known to be
much simpler in these coordinates \cite{conf,GS,PES}.  However, one  
can not
directly use the supercoset approach here since the form fields of
supergravity instead of being constant ($F_{01\dots pr}=1$ in  
$adS\times S$
case) become covariantly constant and depend on transverse coordinates as
$F_{01\dots pI}\sim {y^I \over |y|}$ in cartesian case.  Fortunately, our
approach to the ${G/H}$ supercoset space based on the concept of the  
gauged
superalgebra {\bf G} does not require the structure functions of the
algebra to be true constants.  We are able to solve for the geometry and
the isometries of the supergravity superspace for all cases of the
maximally supersymmetric vacua, characterized by covariantly constant
superfields.
\par
In the discussion section we suggest various possibilities to use our
solution of the superisometries of the $adS\times S$ superspace.  In
particular we comment on BPS states on the worldvolume.  We discuss  
the use
of isometries combined with $\kappa$-symmetry which provides the
superconformal symmetry of the gauge-fixed worldvolume actions including
the Green-Schwarz IIB action in $adS_5\times S^5$ superspace.  Appendix A
contains our conventions with respect to the use of super differential
forms and the gauging of (soft) superalgebras.  Appendix B provides some
elementary useful information on the coset manifolds and superisometries.
\setcounter{equation}{0}
\section{A G-covariant approach to superisometries}
The Killing equations defining the superisometries of the superspace
can e.g.~be found in \cite{Fre}, eq.~(II.3.63). These equations generalize
the  bosonic equations for the isometries of the coset spaces $G/H$,
given in  \cite{Fre} in eq.~(I.6.72) to the supersymmetric case. In both
cases the Killing equations have an $H$-covariant form.
They are particularly difficult to solve to all orders in $\theta$ for
generic cosets $G/H$ including the $adS \times S$ coset superspace.
\subsection{G-covariant Killing equations}
We will take here a different approach to the problem by presenting the
$G$-covariant Killing equations, which will allow us to actually find the
isometries of the $adS\times S$ superspace. We will also show that
our $G$-covariant approach is equivalent to the $H$-covariant one in
\cite{Fre}.
\par
Our starting point is a supergroup $G$ with associated superalgebra {\bf
G}. The generators of the algebra are ${\bf T}_\Lambda$ and satisfy
\begin{equation}{}
[\, {\bf T}_\Lambda , {\bf T}_\Sigma\, \} \equiv
{\bf T}_\Lambda {\bf T}_\Sigma \mp {\bf T}_\Sigma {\bf T}_\Lambda
= f_{\Lambda\Sigma}{}^\Delta {\bf T}_\Delta\,,
\label{Galg}
\end{equation}
where the $+$ sign is taken if both $\Lambda$ and $\Sigma$ are fermionic
indices and $f_{\Lambda\Sigma}{}^\Delta$ are the structure constants.
\par
The standard approach to the supercoset space $G/H$ consists of the
construction of the left-invariant Cartan 1-forms $L$ by
\begin{equation}
\rep (Z)^{-1} d \rep (Z) = L(Z) = L^\Lambda(Z) {\bf T}_\Lambda = dZ^M
L_M{}^\Lambda(Z) {\bf T}_\Lambda\,,
\label{cartandef}
\end{equation}
where $Z^M = \{X^\mu,\theta^{\dot \alpha}\}$ are the supercoset
coordinates
and $\rep (Z)$ is the coset representative.  These forms satisfy the
Maurer-Cartan equations
\begin{eqnarray}
0 = d L - L\wedge L \equiv  \left(d L^\Lambda  + \frac 12 L^\Delta \wedge
L^\Sigma f_{\Sigma\Delta}{}^\Lambda\right) {\bf T}_\Lambda\,.
\label{MC}
\end{eqnarray}
\par
We would like to reinterpret this construction using the concept of
gauging\footnote{The presentation below follows the notation and
conventions of \cite{Toinekar}.} the superalgebra {\bf G}.  To {\sl gauge
the superalgebra} {\bf G} means to introduce the coordinate
dependent gauge field $A$, which is a {\bf G}-valued 1 form,
\begin{equation}
A(Z) \equiv A^\Lambda(Z)  {\bf T}_\Lambda\,
\end{equation}
and to construct a covariant derivative denoted by
\begin{equation}
{\cal D} = d - A\,.
\end{equation}
We define the transformation of the gauge field $A(Z)$ with coordinate
dependent {\bf G}-valued parameter $\Lambda(Z)$, such that in
$\delta(\Lambda) ({\cal D}\Phi)^\alpha$, where $(\Phi)^\alpha$ is a
covariant field, there is no derivative on the parameters
$\Lambda^\Lambda(Z)$.  This can be achieved by taking
\begin{equation}
\delta(\Lambda) A^\Lambda {\bf T}_\Lambda = d\Lambda + [\, A,
\Lambda\,]\nonumber = \left( d\Lambda^\Lambda + \Lambda^\Delta A^\Lambda
f_{\Lambda\Delta}{}^\Lambda\right){\bf T}_\Lambda\,.
\label{gauge}
\end{equation}
One constructs the curvature $F$, which is a {\bf G}-valued 2
form and a covariant field transforming in the adjoint representation,
by
\begin{equation}
{\cal D}^2 = -F\,
\end{equation}
and it follows that
\begin{equation}
F = d A - A\wedge A =
\left(d A^\Lambda + \frac 12 A^\Delta \wedge A^\Sigma
f_{\Sigma\Delta}{}^\Lambda\right) {\bf T}_\Lambda\,.
\end{equation}
\par
The supercoset $G/H$ in the framework of {\sl gauged superalgebra} has
the following properties:
\begin{itemize}
\item The Maurer-Cartan equations associated with the supercoset
construction $G/H$ can be interpreted as the vanishing curvature of
the gauged superalgebra, where the Cartan 1-forms are gauge fields $A$,
which are {\sl pure gauge}
\begin{equation}
L\equiv A \qquad d L - L\wedge L = 0
\qquad \Longrightarrow \qquad F =  d A - A\wedge A  = 0\,.
\end{equation}
\item The gauge transformation with {\bf G}-valued parameter $\Sigma(Z)$
which keeps the gauge field {\sl pure gauge} is defined as the covariant
field $\Sigma(Z)$ with vanishing covariant derivative
\begin{eqnarray}
\delta(\Sigma) A^\Lambda {\bf T}_\Lambda &=& d\Sigma + [\, A, \Sigma\,]=
\left( d\Sigma^\Lambda + \Sigma^\Delta A^\Sigma
f_{\Sigma\Delta}{}^\Lambda\right){\bf T}_\Lambda =0\,.
\label{gauge0}
\end{eqnarray}
\item The integrability condition for the existence of such covariant {\bf
G}-valued field $\Sigma$ with vanishing covariant derivative is the
absence of a curvature, which is equivalent to Maurer-Cartan equations
\begin{eqnarray}
d\Sigma + [\, A, \Sigma\,] =0 \quad \Longrightarrow \quad [F, \Sigma]=0
\quad \Longrightarrow \quad F = d A - A\wedge A = 0\,.
\label{curv0}
\end{eqnarray}
\item A covariantly constant {\bf G}-valued field $\Sigma$ being a gauge
transformation \eqn{gauge}, is by definition $G$-covariant.  We will call
(\ref{gauge0}) for the {\bf G}-valued covariantly constant field
$\Sigma^\Lambda (Z)$ the {\sl G-covariant Killing equations} and the
$\Sigma^\Lambda(Z)$ are referred to as the {\sl G-covariant Killing
superfields}.
\end{itemize}
The relation of our $G$-covariant Killing superfields $\Sigma^\Lambda (Z)$
to the superisometries of the supercoset space $G/H$ is as follows.  The
relevant equation\footnote{In Appendix \ref{app:coset} we present the
derivation of the superisometries for the supercoset space, following
\cite{Fre}.} presents the change of coordinates $\delta Z^M = -\Xi^M $ and
the compensating stability group $H$ transformation with the parameter
$\Lambda^i$ which together keep the Cartan forms invariant:
\begin{eqnarray}
0 &=& \L_\Xi L + d\Lambda + [\, L(Z), \Lambda\,]\nonumber\\
  &=& \left(\L_\Xi L^\Lambda  + d\Lambda^i \delta_i{}^\Lambda + \Lambda^i
L^\Sigma f_{\Sigma i}{}^\Lambda\right) {\bf T}_\Lambda\,. \label{kill2}
\end{eqnarray}
\par
{\sl One can rewrite \eqn{kill2} in $G$-covariant form}.  It turns
out that
this can be done by defining the local parameters
\begin{equation}
\Sigma^{\u M} = \Xi^M L_M{}^{\u M}\,, \qquad \mbox{and}
\qquad \Sigma^i = \Lambda^i + \Xi^M L_M{}^i\,.
\label{connection}
\end{equation}
Using the Maurer Cartan equations (\ref{MC}) we find that \eqn{kill2}
reduces to
\begin{equation}
0 = d \Sigma + [L,\Sigma] = d\Sigma^\Lambda + \Sigma^\Delta L^\Sigma
f_{\Sigma \Delta}{}^\Lambda\,.
\label{Gcovkill}
\end{equation}
This is precisely the $G$-covariant Killing equation which we have derived
in \eqn{curv0}.  It is now clear that if we can solve the covariant
equation we will get the superisometries using (\ref{connection}).
In the same way that \eqn{cartandef} solves the Maurer-Cartan equations
trivially, also the $G$-covariant Killing equation is solved in  
terms of the
coset representative,
\begin{equation}
\Sigma (Z) = \rep^{-1}(Z) \E_0 \rep(Z)\,,
\label{sigmasol}
\end{equation}
where $\E_0$ is a {\bf G}-valued {\sl constant}. In general the Killing
fields depend on all orders of $\theta$.  In the following subsection we
will show how to derive the higher orders in $\theta$ in closed form in
terms of the $\theta$-independent Killing fields, under particular `mild'
assumptions about the coset representative.
\subsection{The G-covariant superisometries to all orders in $\theta$}
First we will show how to solve the equations for Cartan forms to all
orders in $\theta$ reproducing the result from \cite{KRR} but using the
compact notation of the Lie-algebra valued objects and their commutators
defined above.
Consider a boson-fermion split of the algebra {\bf G},
\begin{equation}
{\bf G} = {\bf B} \oplus {\bf F}\,, \label{decompBF}
\end{equation}
where ${\bf B}$ collects the bosonic generators ${\bf B}_A$ and ${\bf F}$
collects the fermionic generators ${\bf F}_\alpha$.  We have to take into
account that the algebra decomposes as
\begin{eqnarray}
{}[{\bf B}, {\bf B}] &=& {\bf B}\,,\nonumber\\
{}[{\bf B}, {\bf F}] &=&  {\bf F}\,,\nonumber\\
\{{\bf F}, {\bf F}\} &=& {\bf B}\,.
\label{BFalg}
\end{eqnarray}
We define the split of a {\bf G}-valued object $A$ into a {\bf B}-valued
and an {\bf F}-valued object according to \eqn{decompBF}
\begin{equation}
A = A^\Lambda {\bf T}_\Lambda = A\bos + A\ferm\,,
\end{equation}
where
\begin{equation}
A\bos = A^A {\bf B}_A\,,\qquad A\ferm = A^\alpha {\bf F}_\alpha.
\end{equation}
The coset representatives are restricted to the form \cite{MT, KRR}
\begin{equation}
\rep(Z) = g(X) e^\Theta\,,
\label{repres}
\end{equation}
where $\Theta$ is an {\bf F}-valued object which may depend on $X$, i.e.
\begin{equation}
\Theta = \Theta\ferm =\Theta^\alpha {\bf F}_\alpha = \theta^{\dot \alpha}
e_{\dot\alpha}{}^\alpha(X) {\bf F}_\alpha \,.
\end{equation}
\par
To get a handle on the higher $\theta$ components of the superfields,
consider the transformation
\begin{equation}
Z^M = \{X^\mu,\theta^{\dot \alpha}\} \rightarrow Z^M_t = \{X^\mu,
t\theta^{\dot\alpha}\} \,, \qquad
\mbox{so}\qquad \Theta \rightarrow t \Theta\,,
\label{Ztrans}
\end{equation}
where $t$ is a real parameter.  Consider the Cartan forms as functions of
the rescaled $\theta$'s
\begin{equation}
L_t = L(X,t\theta) = \rep_t(Z_t)^{-1} d \rep_t(Z_t)\,.
\end{equation}
Differentiating the defining equation \eqn{cartandef} with respect
to  $t$ we
obtain
\begin{equation}
\partial_t L_t = d\Theta + [L_t,\Theta]\,. \label{Gdertcartan}
\end{equation}
We split the Cartan forms as follows
\begin{equation}
L(X,\theta) = L_0(X) + \tilde L(X,\theta)\,
\end{equation}
and make the coset representative has been split such that
\begin{equation}
L_0 = L_0\bos = g^{-1}(X) d g^(X) \qquad \mbox{or}\qquad L_0\ferm = 0\,.
\end{equation}
Therefore the equation \eqn{Gdertcartan} reduces to
\begin{equation}
\partial_t \tilde L_t = D\Theta + [\tilde L_t,\Theta]\,,\label{Gcartan}
\end{equation}
where
\begin{equation}
D\Theta \equiv d\Theta + [L_0, \Theta] = d\Theta + [L_0\bos, \Theta]\,.
\end{equation}
Taking a second derivative of \eqn{Gdertcartan} we obtain
\begin{equation}
\partial_t^2 \tilde L_t = [D\Theta + [\tilde L_t, \Theta],\Theta]\,.
\end{equation}
Using \eqn{decompBF} and some rearrangements of commutators we have
\begin{equation}
\partial_t^2 \tilde L_t\ferm = [\Theta, [\Theta,\tilde L_t\ferm]]
\equiv \M^2 \tilde L_t\ferm\,.
\end{equation}
The matrix $\M^2$ acts on {\bf G}-valued objects $A$ as
\begin{equation}
\M^2 A = [\Theta,[\Theta, A]]\,.
\label{Maction}
\end{equation}
We can solve this oscillator type of equation by giving the initial
conditions
\begin{equation}
\tilde L_{(t=0)}\ferm = 0\,,\qquad (\partial_t \tilde L\ferm)_{(t=0)} = D
\Theta\,.
\end{equation}
The solution reads
\begin{equation}
L_t\ferm = \frac {\sinh t\M}{\M} D\Theta\,.
\end{equation}
Using this result and the initial condition
\begin{equation}
\tilde L_{(t=0)}\bos = 0\,,
\end{equation}
we solve the {\bf B} component of the first order equation
\eqn{Gcartan} to
\begin{equation}
L_t\bos = L_0\bos + 2 \left[\frac{\sinh^2 t\M/2}{\M^2}
D\Theta,\Theta \right]\,.
\end{equation}
The Cartan forms are then obtained by setting $t=1$.
Thus we have reproduced the solution for the Cartan 1-forms
of the superspace \cite{KRR} associated with some superalgebra using the
compact notation of the Lie-algebra valued objects and their commutators.
\par
We use the same procedure to derive the higher order $\theta$  
components of
the Killing superfield $\Sigma(Z)$. Rescaling the coordinates $Z$ again as
in \eqn{Ztrans} we obtain
\begin{equation}
\partial_t \Sigma_t = [\Sigma_t, \Theta]\,,\label{Gparam}
\end{equation}
by differentiating \eqn{sigmasol} w.r.t.~$t$.  We can solve this equations
to all orders in $\theta$ in closed form in terms of the initial  
conditions
\begin{eqnarray}
\Sigma_0\ferm(X) &=& g^{-1}(X) \E\ferm_0 g(X)\,,\qquad
\Sigma_0\bos(X) = g^{-1}(X) \E_0\bos g(X)\,,\nonumber\\
(\partial_t \Sigma\ferm)_{(t=0)} &=& [\Sigma_0\bos(X), \Theta]
\equiv {\cal B} \Theta\,.
\label{initsigma}
\end{eqnarray}
Taking a second derivative of \eqn{Gparam} we get
\begin{equation}
\partial_t^2 \Sigma_t = [[\Sigma_t,\Theta],\Theta]\,,
\end{equation}
For the fermionic direction using \eqn{decompBF} this gives again an
oscillator type equation.  After rearranging the commutators we get
\begin{equation}
\partial_t^2 \Sigma_t\ferm = \M^2 \Sigma_t\ferm\,,
\label{oscferm}
\end{equation}
with $\M$ given in \eqn{Maction}.
The solution to this equation reads, with initial conditions
\eqn{initsigma}
\begin{equation}
\Sigma_t\ferm = \cosh t\M \Sigma_0\ferm + \left(\frac {\sinh
t\M}{\M}\right)
{\cal B}\Theta\,.
\end{equation}
Taking this solution into the {\bf B} component of \eqn{Gparam} leads to
\begin{equation}
\Sigma_t\bos = \Sigma_0\bos + \left[\frac {\sinh
t\M}{\M}\Sigma\ferm,\Theta\right] + 2 \left[\frac {\sinh^2
t\M/2}{\M^2} [L_0,\Theta], \Theta \right]\,.
\label{bos}
\end{equation}
The $G$-covariant Killing superfields are recovered at $t=1$ in terms of
the known $\theta$-independent Killing fields $\Sigma_0^\Lambda(X)$.  \par
To conclude let us summarize what we have now.  Given any supercoset space
$G/H$ and the coset representative (i.e.~choice of coordinates) of the
bosonic subspace $g(X)$, we can construct the complete geometric
superfields $L(Z)$ and Killing superfields $\Sigma(Z)$.  By taking the
expression \eqn{repres} for the supercoset representative we have assumed
that fermionic Cartan 1-forms vanish at vanishing $\theta$, i.e.
$L\ferm_0=0$.  The usual Killing vectors, Killing spinors and compensating
stability group transformations can be recovered by identifying the coset
generators and stability group generators and use the transformation
\eqn{connection}.  We summerize here the relevant expressions for the
Cartan 1-forms and $G$-covariant Killing fields in index notation, i.e.~in
terms of the structure constants of the {\bf G} algebra,
\begin{eqnarray}
L^\alpha &=& \left(\frac {\sinh \M}{\M} D\Theta\right)^\alpha\,,\qquad
L^A = L_0^A + 2 \Theta^\alpha f_{\alpha\beta}^A\left(\frac{\sinh^2
\M/2}{\M^2} \Theta \right)^\beta\,,\nonumber\\
\Sigma^\alpha &=& ( \cosh t{\cal M} \; \Sigma_0)^\alpha + \left(
\frac{\sinh {\cal M}}{{\cal M}} {\cal B} \Theta
\right)^\alpha\,,\nonumber\\
\Sigma^A &=& \Sigma^A_0 + \Theta^\alpha f_{\alpha \beta}^A \left( \frac
{\sinh{\cal M}}{{\cal M}} \; \Sigma_0 + 2\frac {\sinh^2 {\cal M}/2}{{\cal
M}^2} {\cal B}\Theta \right )^\beta\,,\nonumber\\
({\cal B}\Theta)^\alpha &=& \Theta^\beta \Sigma_0^A f_{A\beta}^\alpha\,,
\nonumber\\ (\M^2)^\alpha{}_\beta &=& f^\alpha_{A\gamma} \Theta^\gamma
\Theta^\delta f_{\delta\beta}^A\,.
\end{eqnarray}
The most difficult point in establishing a nice superspace and its
symmetries is the choice of the coset representative for the bosonic
subspace, i.e. choosing appropriate coordinates.
In particular applications one could prefer having an explicit expression
for the more `familiar' $H$-covariant super Killing vector and spinor and
the comensating $H$-transformation.  We will derive these expressions in
the next section in a special gauge.
\setcounter{equation}{0}
\section{H-covariant superisometries in Killing-spinor\\ gauge}
In this section we will recover the superisometries in $H$-covariant form
which gives the transformation of the coordinates
\begin{equation}
\delta Z^M = - \Xi^M\,
\end{equation}
and the corresponding compensating stability group transformation given by
the parameter $\Lambda^i$ in \eqn{kill2}.
\par
We consider the coset decomposition of the algebra
\begin{equation}
{\bf G} = {\bf K} \oplus {\bf H}\,, \label{decompKH}
\end{equation}
where {\bf K} collects the ``coset generators'' ${\bf K}_{\u M}$ and {\bf
H} collects the stability algebra generators ${\bf H}_i$.  The algebra
decomposes into
\begin{eqnarray}
{}[{\bf K}, {\bf K}\} &=& {\bf K} \oplus {\bf H}\,,\nonumber\\
{}[{\bf H}, {\bf K}\} &=& {\bf K}\,,\nonumber\\
{}[{\bf H}, {\bf H}\} &=& {\bf H}\,.  \end{eqnarray}
In the flat and near-horizon superspaces we deal with a {\sl
homogeneous} superspace which is not symmetric\footnote{We thank P.~Howe
for pointing this out.}.  A symmetric space would have ${}[{\bf K}, {\bf
K}\} = {\bf H}$.  And we will consider only a {\sl reductive}
decomposition, ${}[{\bf H}, {\bf K}\} = {\bf K}$, which is satisfied  
by the
flat and near-horizon superspaces.  A more general form of the coset would
also contain ${}[{\bf H}, {\bf K}\} = {\bf K} \oplus {\bf H}$.
We split the Cartan 1-forms $L$ and $G$-covariant parameters $\Sigma$
according to \eqn{decompKH}
\begin{eqnarray} L &=& E + \Omega = E^{\u M} {\bf K}_{\u M} +  
\Omega^i {\bf
K}_i\,,\nonumber\\ \Sigma &=& \hat \Xi + \hat \Lambda = \hat \Xi^{\u M}
{\bf K}_{\u M} + \hat \Lambda^i {\bf H}_i\,.
\end{eqnarray}
\par
Since in this paper we are interested in maximally supersymmetric
superspaces we will restrict ourselves to the case where
\begin{equation}
{\bf F} \subset {\bf K}\qquad\mbox{or}\qquad {\bf F} \cap
{\bf H} = 0\,.
\end{equation}
The $G$-covariant parameters $\{\hat \Xi^{\u M}, \hat \Lambda^i\}$ are
related to the $H$-covariant ones through \eqn{connection},
\begin{equation}
\hat \Xi^{\u M} = \Xi^M E_M{}^{\u M}\,,\qquad \hat \Lambda^i = \Lambda^i +
\Xi^M \Omega_{M}{}^i\,.
\end{equation}
We can split the bosonic generators of the previous section into
\begin{equation}
{\bf B} = \{{\bf P}_a , {\bf M}_i\}\,, \qquad P_a \in {\bf
K}\ \mbox{and}\ M_i \in {\bf H}\,.
\end{equation}
Before we write down the full $\theta$-dependent parameters, we repeat the
$\theta=0$ killing equations.  We denote the  Cartan 1-forms
$L_0^\Lambda$ by
\begin{eqnarray}
&&\psi \equiv L_0^\alpha{\bf F}_\alpha = \psi^\alpha{\bf F}_\alpha =  
dx^\mu
\psi_\mu^\alpha {\bf F}_\alpha\,,\nonumber\\
&&e \equiv L_0^a {\bf P}_a = e^a {\bf P}_a = dx^\mu e_\mu{}^a{\bf
P}_a\,,\nonumber\\
&&\omega \equiv L_0^i {\bf H}_i = \omega^i {\bf H}_i = dx^\mu \omega^i_\mu
{\bf H}_i\,,
\end{eqnarray}
where in the construction outlined above $\psi$ has been chosen to vanish.
The names have not been chosen at will since they are the vielbein $e^a$,
the gravitino $\psi^a$ and the spin connection $\omega^i$ of the geometry,
which is a solution to a supergravity theory. We will clarify the relation
between supergravity solutions and supercoset spaces in later sections. We
introduce also the more familiar parameters
\begin{eqnarray}
&&\epsilon\equiv \hat \Xi_0^\alpha {\bf F}_\alpha =
\epsilon^\alpha{\bf F}_\alpha\,,\nonumber\\
&&\hat \xi \equiv \hat \Xi_0^a {\bf P}_a = \xi^a {\bf P}_a= \xi^\mu
e_\mu{}^a{\bf P}_a\,,\nonumber\\
&&\hat \ell \equiv \hat \Lambda^i_0 {\bf M}_i = \hat \ell^i{\bf M}_i =
(\ell^i + \xi^\mu \omega_\mu{}^i){\bf M}_i\,.
\label{GtoHparam}
\end{eqnarray}
The $\theta$-independent part of \eqn{Gcovkill} splits up into
($\psi = 0$)
\begin{eqnarray} 0 &=& (d \epsilon + [\omega, \epsilon] + [e
,\epsilon])^\alpha {\bf F}_\alpha\,,\nonumber\\ 0 &=& (d \xi + [\omega,
\hat \xi] + [e,\hat \ell] + [e,\hat \xi])^a {\bf P}_a\,,\nonumber\\
0 &=& (d \hat \ell + [\omega, \hat \ell] + [\omega,\hat \xi] + [e,\hat
\xi])^i{\bf M}_i\,.
\end{eqnarray}
We rewrite the last two equations using \eqn{GtoHparam} and obtain the set
of killing-equations
\begin{eqnarray}
0 &=& \delta \psi^\alpha {\bf F}_\alpha = (d \epsilon + [\omega, \epsilon]
+ [e ,\epsilon])^\alpha {\bf F}_\alpha\,,\label{killferm}\\ 0 &=& \delta
e^a {\bf P}_a = (\L_\xi e^a + [e,\ell]^a){\bf P}_a\,,\\ 0 &=& \delta
\omega^i {\bf M}_i = (\L_\xi \omega^i + d\ell^i + [\omega,\ell]^i){\bf
M}_i\,.
\end{eqnarray}
These equations give the non-vanishing values of Killing
spinors $\epsilon^\alpha (X) $, Killing vectors $\xi^\nu(X)$ and
compensating Lorentz transformation $ \ell^i(X)$ at order $\theta^0$.
{}From the supergravity point of view the meaning of these equations  
is that
this set of transformations leaves the supergravity solution invariant,
i.e. under these supersymmetry transformations, general coordinate
transformations and Lorentz transformations,
the gravitino, vielbeins and spin connections, do not change.
\par
Before we go on to derive $\Xi^M$ and $\Lambda^i$ to all orders in
$\theta$, we will take a special value of $e_{\dot \alpha}{}^\alpha$
related to the solution of the killing spinor equation \eqn{killferm}.  It
is known that the solution to \eqn{killferm} can be written as
\begin{equation}
\epsilon^\alpha (X) = \epsilon_0^{\dot\alpha} {\cal
K}(X)_{\dot \alpha}{}^\alpha\,, \label{Kmatrix}
\end{equation}
where $\epsilon_0$ is a constant spinor.  As the matrix $e_{\dot
\alpha}{}^\alpha$ was left unspecified until now we can still {\sl choose}
it to be equal to ${\cal K}$, called the {Killing-spinor gauge}. Doing so
we find
\begin{equation}
e_{\dot \alpha}{}^\alpha = {\cal K}_{\dot \alpha}{}^\alpha\quad
\Longrightarrow\quad (D\Theta) = (d\theta)^{\dot \alpha} {\cal
K}_{\dot\alpha}{}^\alpha{\bf F}_\alpha\,.
\end{equation}
This leads to simplifications of the Cartan 1-forms
\begin{equation}
L_\mu{}^\Lambda (X,\theta) = L_0{}_\mu{}^\Lambda(X)\,.
\end{equation}
In this gauge we obtain
\begin{eqnarray} \Xi^{\alpha} {\bf F}_\alpha &\equiv& \Xi^{\dot\alpha}
{\cal K}_{\dot\alpha}{}^\alpha {\bf F}_\alpha = (\M \coth\M \epsilon) +
({\cal B}\Theta) \,,\nonumber\\
\Xi^a {\bf P}_a &\equiv& \Xi^\mu e_\mu{}^a {\bf P}_a = \xi^a {\bf P}_a +
\left[\frac {\tanh\M/2}{\M} \epsilon,\Theta\right]^a {\bf P}_a
\,,\nonumber\\
\Lambda^i &=& \ell^i{\bf M}_i + \left[\frac {\tanh\M/2}{\M}
\epsilon,\Theta\right]^a e_a{}^\mu \omega_\mu{}^i {\bf M}_i + \left[\frac
{\tanh\M/2}{\M} \epsilon,\Theta\right]^i{\bf M}_i\,.
\end{eqnarray}
This concludes the derivation of the superisometries.  On the coordinates
they act as
\begin{equation}
\delta X^\mu = - \Xi^\mu(X,\theta)\,, \qquad
\delta \theta^{\dot \alpha} = -\Xi^{\dot\alpha} (X,\theta)\,.
\end{equation}
\setcounter{equation}{0}
\section{Superisometries of maximally supersymmetrc vacua}
In this section we will present the superisometries for the two maximally
supersymmetric vacua, i.e. flat space, where G is the super  
Poincar\'e group
and $adS_{p+2} \times S^{d-p-2}$ with $G$ the relevant extended
superconformal group.

\subsection{Flat superspace}
As a check on the general results obtained above we apply our formulae to
the flat superspace. The coset starts with the super
Poincar\'e group $G$.  The algebra is given by
\begin{eqnarray}
{}[M_{ab}, M_{cd}] &=& \eta_{a[c} M_{d]b} - \eta_{b[c}
M_{d]a}\,,\nonumber\\ {}[P_a, M_{bc}] &=& \eta_{a[b} P_{c]}\,,\nonumber\\
{}[M_{ab}, Q_\alpha] &=& -\ft14 (\gamma_{ab} Q)_\alpha\,,\nonumber\\
{}\{Q_\alpha, Q_\beta\} &=& (\gamma^a)_{\alpha\beta} P_a\,.
\end{eqnarray}
We make the following split
\begin{equation} {\bf H} = \{ M_{ab} \}\,,\qquad \mbox{and}\qquad  
{\bf K} =
\{P_a, Q_\alpha\}\,.
\end{equation}
The indices of the previous sections are thus
\begin{equation} \Lambda = \{a, (ab), \alpha \}\,,\qquad A=\{a,
(ab)\}\,,\qquad i=\{(ab)\}\,, \qquad{\u M} = \{a, \alpha\}\,.
\end{equation}
The space-time fields are given by
\begin{equation}
e_\mu^a = \delta_\mu^a\,,\qquad \psi_\mu = 0\,,\qquad \omega^{ab}_\mu =
0\,,
\end{equation}
and the solutions to the space-time killing equations is
\begin{equation}
\xi^\mu = a^\mu + \lambda_M^{\mu\nu} x_\nu\,,\qquad \epsilon^\alpha(x) =
\epsilon_0^\alpha\,,\qquad \ell^{ab} = \lambda_M^{\mu\nu}
\delta_\mu^a\delta_\nu^b\,,
\end{equation}
where $a^\mu, \lambda_M^{\mu\nu}$ and $\epsilon^\alpha_0$ are constant
parameters.  The matrix ${\cal K}_{\dot \alpha}{}^{\beta} = \delta_{\dot
\alpha}^\beta$ and the matrix $\M$ vanishes.  The Cartan 1-forms are given
by
\begin{equation}
E^\alpha = d\theta^\alpha\,,\qquad E^a = dx^a + \bar\theta \gamma^a
d\theta\,.
\end{equation}
We read off that
\begin{eqnarray}
\Xi^\mu &=& a^\mu + \lambda_M^{\mu\nu} x_\nu + \bar \theta \gamma^\mu
\epsilon_0\,,\nonumber\\
\Xi^\alpha &=& \epsilon_0^\alpha + \ft14 (\lambda_M \cdot \gamma
\theta)^\alpha\,,
\end{eqnarray}
which gives the well-known  superspace and superisometries.
\subsection{adS$_{p+2}\times\,$S$^{d-p-2}$ superspace}
To construct this superspace we start from the superconformal group $G$
which has $SO(p+1,2) \times SO(d-p-1)$ as its bosonic subgroup.  For this
supercoset the stability group $H$ is the product group $SO(p+1,1)\times
SO(d-p-2)$, which is purely bosonic.  In particular we will derive  
explicit
expressions for the Killing vectors $\xi^\mu$ and Killing spinors
$\epsilon^\alpha$ as well as the compensating $H$-transformation  
parameters
$\ell^{ab}$ for the prototypical brane solutions to various supergravity
theories which have a near-horizon geometry of the form $adS_{p+2} \times
S^{d-p-2}$ and are summarized in table \ref{tab:sol}.
\begin{table}
\begin{center}
\begin{tabular}{|l|l|c|} \hline
supergravity & brane solution & supergroup $G$\\\hline\hline
$d=11$ sugra& $M2$ brane $(p=2)$& $OSp(8|4)$\\
$d=11$ sugra& $M5$ brane $(p=5)$& $OSp(6,2|4)$\\
$d=10$ IIB sugra & $D3$ brane $(p=3)$ & $SU(2,2|4)$\\
$d=6$ $(2,0)$ sugra & self-dual string $(p=1)$ & $SU(1,1|2)^2$\\
$d=4$ $N=2$ sugra & Reissner-Nordstr\o m black hole $(p=0)$ &
$SU(1,1|2)$\\\hline
\end{tabular}
\end{center}
\caption{Supergravity brane solutions with $adS_{p+2}\times S^{d-p-2}$.
\label{tab:sol}}
\end{table}
The last two solutions in this table can be obtained as intersecting brane
solutions.  \par The bosonic isometries, which form the group $SO(p+1,2)
\times SO(d-p-1)$, can be treated in a uniform way for all cases in table
\ref{tab:sol}.  The geometry is described by the product space metric
\cite{conf}
\begin{equation}
ds^2 = dx^\mu g_{\mu\nu} dx^\nu = g^{adS} + g^{S}\,,
\end{equation}
where
\begin{eqnarray}
g^{adS}&=& \left(\frac rR\right)^{2/w} dx^m \eta_{mn} dx^n
+ \left(\frac Rr\right)^2 dr^2\,,\nonumber\\
g^S &=& R^2 d\Omega^2\,.
\label{metric}
\end{eqnarray}
Besides the metric there are also the non-trivial forms which will be
introduced in due course and are proportional to the volume forms of the
$adS_{p+2}$ and $S^{d-p-2}$ spaces.  The coordinates $X^\mu$ are  
split into
$adS$ coordinates $X^{\hat m} = \{x^m,r\}$, $m=0,\dots,p$ and coordinates
on the sphere, here we take angular coordinates $X^{m'} = \{\phi^{m'}\}$,
$m'=1,\dots,d-p-2$.  $d\Omega^2$ is the metric on the unit sphere.  The
number $w={p+1\over d-p-3}$ has a fixed value that gives the ratio between
the radius of the $adS$ space and the sphere $S$ for the specific
supergravity solutions.
\par
The Killing vectorfield $\xi\equiv \xi^\mu \partial_\mu$ is the  
solution to
\begin{equation}
\L_{\xi} g = 0\,.
\end{equation}
Due to the fact that we are dealing with a product geometry, in the
coordinates given above this equation splits into
\begin{equation}
\L_{\xi_{adS}} g^{adS} = 0\,,\qquad \L_{\xi_{S}} g^S = 0\,.
\end{equation}
The components of the killing-vector field $\xi_{adS}$ of the $adS_{p+2}$
space are given by \cite{conf}
\begin{eqnarray}
\xi^{m}_{adS} &=& \xi_{conf}^m(x) + (wR)^2 \left(\frac
Rr\right)^{2/w}\Lambda_K^m\nonumber\\
&=& a^m + \lambda^{mn}_M x_n + \lambda_D x^n + (x^2 \Lambda_K^m - 2 x^m
x\cdot \Lambda_K) + (wR)^2 \left(\frac
Rr\right)^{2/w}\Lambda_K^m\,,\nonumber\\
\xi^{r}_{adS} &=& - w \Lambda_D(x) r\,,
\label{killingvectorsadS}
\end{eqnarray}
where
\begin{equation}
\Lambda_D(x) = \ft 1{p+1} \partial_m \xi^m_{conf} = \lambda_D - 2 x\cdot
\Lambda_K\,.
\end{equation}
The compensating transformation of the stability subgroup $SO(p+1,1)$ is
\begin{eqnarray}
\ell^{\u m\u n} &=& \Lambda^{mn}_M(x)\, \delta_m^{\u m}\delta_n^{\u
n}\,,\nonumber\\
\ell^{\u m \u r} &=& -2 (wR) \left(\frac Rr\right)^{1/w} \Lambda_K^m\,
\delta_m^{\u m}\,,
\label{compadS}
\end{eqnarray}
where
\begin{equation}
\Lambda^{mn}_M(x) = -2 \partial^{[m} \xi_{conf}^{n]} = \lambda_M^{mn} - 4
x^{[m} \Lambda_K^{n]}\,.
\end{equation}
This is obtained as a solution to
\begin{equation}
0 = \L_{\xi_{adS}} e^{\u {\hat m}} - \ell^{\u{\hat m}\u {\hat n}}
e_{\u{\hat n}}\,.
\end{equation}
In these Killing vectors and compensating transformations, we identify the
conformal Killing vectors $\xi_{conf}^m$ and conformal `compensating'
transformations with parameters $\Lambda_D(x)$ and $\Lambda_M^{mn}(x)$.
\par On $S^{d-p-2}$ the killing vectors $\xi_{S}^{m'}$ as well as the
compensating transformation $\ell^{\u m'\u n'}$ from the stability  
subgroup
$SO(d-p-2)$ are somewhat complicated functions of the angles.  These
transformations can be expressed in terms of the R-symmetry subgroup  
of the
superconformal group which is the symmetry of the surface
\begin{equation}
X^{\hat m'} \delta_{\hat m' \hat n'} X^{\hat m'} = R^2
\end{equation}
where $\hat m'$ takes values $1, \dots , d-p-1$.  The $SO(d-p-1)$
invariance is linearly realized on the $X^{\hat m'}$ coordinates,
\begin{equation}
\delta X^{\hat m'} = -\Lambda^{\hat m'}{}_{\hat n'} X^{\hat n'}\,;\qquad
\Lambda_{\hat m'\hat n'} = -\Lambda_{\hat n'\hat m'}\,.
\end{equation}
This surface condition can be solved in terms of hyperspherical  
coordinates
$\phi^{m'}$
\begin{equation}
X^{\hat m'} = \left\{ R \cos \phi^1\,,
\dots\,, R\cos \phi^{d-p-2}\prod_{i=1}^{d-p-3}\sin \phi^i\,, R
\prod_{i=1}^{d-p-2} \sin \phi^i \right\}\,,
\end{equation}
with
\begin{eqnarray}
\phi^{m'} &=& \arctan \left[ \frac {\left(\sum_{k'=m'+1}^{d-p-1}
(X^{k'})^2\right)^{1/2}}{X^{m'}}\right]\,.
\label{angles}
\end{eqnarray}
The change of angular variables $\delta \phi^{m'}= -\xi_S^{m'}$ which
preserves the metric on the sphere can be deduced from (\ref{angles}).
These transformations do not preserve the vielbein forms and one has to
find also the compensating transformation which preserves the vielbeins.
It is important that all R-symmetry transformations depend only on angles
and global parameters $\Lambda^{\hat m'}{}_{\hat n'}$.  As an example
consider here the case of $S^2$ with two angles, $d\Omega^2= (d\phi^1)^2 +
(\sin \phi^1d\phi^2)^2$, i.e.  $m'=1,2$ and $\hat m'=1,2,3$.  One finds
that the $SO(3)$ isometry (R-symmetry) is realized on the two angles  
of the
sphere as
\begin{eqnarray}
\xi^{\phi_1} &=& -\cos \phi_2 \Lambda^{12} - \sin \phi_2
\Lambda^{13}\,,\nonumber\\
\xi^{\phi_2} &=& \cot \phi_1 \sin\phi_2 \Lambda^{12} - \cot \phi_1
\cos\phi_2 \Lambda^{13} - \Lambda^{23}\,,
\label{killingvectorssphere}
\end{eqnarray}
and the compensating $SO(2)$ transformation is \begin{eqnarray}
\ell^{\u\phi_1\u\phi_2} = \sin^{-1} \phi_1 (\sin \phi_2 \Lambda^{12}  
- \cos
\phi_2 \Lambda^{13})\,.
\label{compsphere}
\end{eqnarray}
It is not so difficult to see that $\frac\partial{\partial \phi^{d-p-2}}$
will always be a Killing vector and therefore there will always be
conservation of momentum in this angular direction.  This concludes the
derivation of the bosonic isometries.
\par
The Killing spinors cannot be treated in the same uniform way because
properties of spinors depend on the dimension and signature.  However, a
detailed study of the relevant Killing spinor equations (vanishing
transformation of fermions at vanishing fermions in the supergravity
theory) for each case in table \ref{tab:sol} reveals a general  
structure of
the Killing spinors.  The difference between all cases is in the split of
the generic fermions by some specific projection operator.  First we will
give the supergravity Killing-spinor equations.
\leftmargini=5mm
\leftmarginii=2mm
\begin{enumerate}
\item {\sl The $G=OSp(M|N)$ cases}
\begin{enumerate}
\item {\sl $M2$ and $M5$ brane solutions to $d=11$ supergravity}\\
Besides the metric \eqn{metric}, the near-horizon $M2$ and $M5$ solutions
have a non-vanishing 4-form field strength and its dual 7-form field
strength, given by
\begin{eqnarray}
&\mbox{M2}&\qquad F^4_{\u {\hat m_1}\dots\u{\hat m}_4} = -\frac 6R
\epsilon_{\u {\hat m}_1\dots \u {\hat m}_4}\qquad F^7_{\u m'_1\dots \u
m'_7} = \frac 6R \epsilon_{\u m'_1\dots \u m'_7}\,,\nonumber\\
&\mbox{M5}&\qquad F^7_{\u {\hat m}_1\dots \u {\hat m}_7} = \frac 3R
\epsilon_{\u {\hat m}_1\dots \u {\hat m}_7}\,,\qquad F^4_{\u m'_1\dots \u
m'_4} = \frac 3R \epsilon_{\u m'_1\dots \u m'_4}\,.
\label{Mfieldstrengths}
\end{eqnarray}
The Killing spinor equation is given by the vanishing variation of the
gravitino at vanishing gravitino \cite{CJS}
\begin{equation} 0 = \delta \psi_\mu = \partial_\mu \epsilon + \frac 14
\omega_\mu^{ab} \Gamma_{ab}\epsilon + \frac 1{288}e_\mu{}^a
\left(\Gamma_a{}^{bcde} - 8 \delta_a^{[b} \Gamma^{cde]}\right)
F_{bcde}\epsilon\,.
\end{equation}
The spinor $\epsilon$ is a 32 component Majorana spinor.  To solve these
equations it turns out that it is useful to introduce projections of the
spinors
\begin{equation}
\epsilon_\pm = {\cal P}_\pm \epsilon = \ft12 (1\pm \Gamma^{01\dots
p})\epsilon\,.
\label{OSPproj}
\end{equation}
\end{enumerate}
\item {\sl The $G= SU(M|N)$ cases}
\begin{enumerate}
\item{\sl D3 brane solution to $d=10$ IIB supergravity}\\
The $D3$ brane solution to 10 dimensional IIB supergravity is given by the
metric \eqn{metric} with $p=3$, $w=1$, the self dual 5-form with  
components
\begin{equation}
F_{\u{\hat m}_1 \dots \u {\hat m}_5} = \frac 4R \epsilon_{\u{\hat m}_1
\dots \u {\hat m}_5} \,,\qquad F_{\u m'_1\dots \u m'_5} = \frac 4R
\epsilon_{\u m'_1\dots \u m'_5}\,,
\end{equation}
and a {\sl constant} dilaton $\phi$.  The fermions and all other forms
vanish.  \par With this solution the Killing spinor equations  
\cite{HW} can
be cast in the form
\begin{eqnarray}
&&0 = \delta \lambda = 0\,,\nonumber\\
&&0 = \delta \psi = \partial_\mu \epsilon + \frac14 \omega_\mu{}^{ab}
\Gamma_{ab}\epsilon - \frac i{16\cdot 5!} e_\mu{}^a F^+_{bcdef}
\Gamma^{bcdef} \Gamma_a \epsilon\,,
\end{eqnarray}
where the dilatino
$\lambda$ is a complex are left-handed Weyl spinors (or two Majorana-Weyl
spinors), i.e.  $\Gamma_{11}\lambda = \lambda$.  The parameters
$\epsilon^i$ and the gravitini $\psi_\mu^i$ are right-handed Weyl spinors,
$\Gamma_{11} \epsilon = -\epsilon$ and $\Gamma_{11} \psi_\mu = -\psi_\mu$.
The variation of the dilatino vanishes trivially for vanishing fermions
since it transforms into the derivative of the dilaton and it does not
transform into the self-dual 5-form.  To solve the Killing spinor equation
we introduce
\begin{equation}
\epsilon_\pm = {\cal P}_\pm \epsilon = \ft12 (\epsilon \pm i\Gamma^{0123}
\epsilon)\,,
\label{D3proj}
\end{equation}
\item{\sl self-dual string solution to $d=6$ $(0,2)$ supergravity}\\
The chiral $d=6$ $(0,2)$ pure supergravity theory contains \cite{6dim} the
graviton $g_{\mu\nu}$, 4 chiral gravitini $\psi^i_\mu$ and 5 self-dual
tensors $B^{a'}_{\mu\nu}$, where $a'$ is an $SO(5)$ vector index.  This
theory has also been known as pure $N=4b$ supergravity.  The self-dual
string solution to this theory has the metric \eqn{metric} with $p=1$,
$w=1$, the self dual 3-form with components
\begin{equation}
F^{a'}_{\u{\hat m}_1 \u{\hat m}_2 \u{\hat m}_3} = \frac 2R
\epsilon_{\u{\hat m}_1 \u{\hat m}_2 \u{\hat m}_3} \delta^{a'}_{1'}
\,,\qquad F^{a'}_{\u m'_1 \u m'_2 \u m'_3} = \frac 2R \epsilon_{\u m'_1 \u
m'_2 \u m'_3}\delta^{a'}_{1'}\,
\end{equation}
and the gravitini vanish.
\par
With this solution the vanishing variation of the gravitini are
\cite{6dim}
\begin{equation} 0 = \delta \psi_\mu^i = \partial_\mu
\epsilon^i + \frac 14 \omega_\mu^{ab} \Gamma_{ab}\epsilon^i - \frac 1{8}
e_\mu{}^a F^1_{abc} \Gamma^{bc} (\Gamma')^i{}_j \epsilon^j\,.
\end{equation}
Here $\epsilon^i$ (and also $\psi_\mu^i$) is a righthanded
($\Gamma_7 \epsilon^i = -\epsilon^i$) symplectic Majorana-Weyl  
spinor.  The
$USp(4)$ index $i$ is raised and lowered by $\epsilon^i = \Omega^{ij}
\epsilon_j$ and $\epsilon_i = \epsilon^j \Omega_{ji}$, where $\Omega_{ij}$
is the antisymmetric $USp(4)$ metric and we define $\Omega^{ik}  
\Omega_{jk}
=\delta^i{}_j$.  The matrix $\Gamma'\equiv \Gamma'_{1'}$ is an $SO(5)$
$\Gamma$-matrix satisfying $(\Gamma')^2 = 1$ and $[\Gamma^a, \Gamma']=0$.
For the self-dual string solution we introduce the projections
\begin{equation} \epsilon_\pm \equiv \epsilon^i_\pm = {\cal P}_\pm
\epsilon^i = \ft12 (\delta^i{}_j \pm \Gamma^{01} (\Gamma')^i{}_j)
\epsilon^j\,.
\label{SDSproj}
\end{equation}
\item{\sl Reissner-Nordstr\o m black hole solution to $d=4$ $N=2$
supergravity}\\
The metric \eqn{metric} for $p=0$, $w=1$ and $d=4$ is the near-horizon
geometry of an extremal Reissner-Nordstr\o m black hole solution in
isotropic coordinates, where $R$ is proportional to the mass of the black
hole .  This solution can be obtained as a 1/2 BPS solution to $d=4$ $N=2$
pure supergravity.  We will restrict to the case where the black hole is
electrically charged and therefore we restrict to
the case that the graviphoton field strength has components
\begin{equation}
F_{\u {\hat m}_1\u {\hat m}_2} = - \frac 1R \epsilon_{\u{\hat m}_1\u {\hat
m}_2}\,.
\end{equation}
\par
The Killing-spinor equation is
\begin{equation}
0=\delta \psi_\mu^i = \partial_\mu \epsilon^i + \frac 14 \omega_\mu^{ab}
\Gamma_{ab} \epsilon^i + e_\mu{}^a F^-_{ab} \Gamma^b \varepsilon^{ij}
\epsilon_j\,, \end{equation} where $\epsilon^i$ (and also $\psi^i_\mu$) is
a generic doublet ($i=1,2$) of Majorana spinors satisfying $-i
(\epsilon_i)^\dagger \Gamma_0 = (\epsilon^i)^T {\cal C}$ with ${\cal C}$
the 4 dimensional charge conjugation matrix.  The place of the index $i$
denotes the chirality of the spinor, e.g.~$\Gamma_5 \epsilon^i =
\epsilon^i$ and $\Gamma_5 \epsilon_i = -\epsilon_i$.  $F^-_{ab}$ is the
anti-self dual part of $F_{ab}$.  To solve the Killing-spinor equation for
this case we introduce \begin{equation} \epsilon_\pm \equiv \epsilon_\pm^i
= {\cal P}_\pm \epsilon^i = \ft12 (\epsilon^i \pm \Gamma^{0}
\varepsilon^{ij} \epsilon_j)\,.
\label{BHproj}
\end{equation}
\end{enumerate}
\end{enumerate}
Using the appropriate projector for each case at hand we can rewrite the
killing spinor equations as
\begin{eqnarray}
0&=& \partial_m \epsilon_+ + \frac 1{wR} \left(\frac rR\right)^{1/w}
\Gamma_m \Gamma_r \epsilon_-\,,\\
0&=& \partial_m \epsilon_-\,,\\ 0&=& \partial_r \epsilon_{\pm} \mp \frac
1{2w} \frac 1r \epsilon_{\pm}\,,\\
0&=& \nabla_{m'} \epsilon_{\pm} \mp \frac 1{2R} \Gamma_r \tilde
e_{m'}{}^{\u m'} \Gamma_{m'} \epsilon_{\pm}\,,
\label{spherekill}
\end{eqnarray}
where $\nabla_{m'}$ is the covariant derivative and $\tilde e_{m'}{}^{\u
m'}$ is the vielbein of $S^{d-p-2}$.  The explicit solutions to the
Killing-spinor equations \eqn{spherekill} on the sphere have been derived
in \cite{LPR}.  The full solution to the Killing-spinor equations is
($\epsilon = \epsilon_+ + \epsilon_-$)
\begin{eqnarray} \epsilon_- &=& - (wR) \left(\frac Rr\right)^{1/2w}
\Gamma_r u(\phi^{m'}) \eta \,,\nonumber\\
\epsilon_+ &=& \left(\frac rR\right)^{1/2w} u(\phi^{m'})  
\left(\epsilon_0 +
{x \cdot \Gamma} \eta\right)\,,
\label{KS}
\end{eqnarray}
where $\epsilon_0$ and $\eta$ are constant spinors, satisfying
\begin{equation}
\epsilon_0 = {\cal P}_+\epsilon_0\,,\qquad \eta = \Gamma_r
{\cal P}_- \Gamma_r \eta\,.
\label{projections}
\end{equation}
The function $u (\phi^{m'})$ results from the Killing-spinor equation on
the sphere and are given by \cite{LPR}
\begin{equation}
u (\phi^{m'}) = (\cos \ft {\phi^1}2 + \Gamma_{r1'} \sin \ft
{\phi^1}2)\prod_{k'=2}^{d-p-2} (\cos \ft {\phi^{k'}}2 + \Gamma_{(k'-1)k'}
\sin \ft {\phi^{k'}}2)\,.
\end{equation}
Interestingly enough again the superconformal Killing spinor in Minkowski
flat background
\begin{equation}
\epsilon(x) = (\epsilon + x\cdot \Gamma \eta)\,,
\end{equation}
where $\epsilon$ is the parameter of rigid supersymmetry and $\eta$ is the
parameter of rigid conformal supersymmetry, appears in this form of the
Killing spinors.  This Killing spinor was already introduced in the
pioneering papers on supersymmetry by Wess and Zumino \cite{WZ}.  To
conclude this derivation we note that we did not have to choose a
particular split of the $\Gamma$-matrices in terms of tensor products of
lower dimensional $\Gamma$-matrices.  Also the Killing spinors are  
given by
projections of the generic spinor of the background supergravity theory.
\par
Now we are ready to present the superisometries of $adS_{p+2} \times
S^{d-p-2}$ superspace in the Killing spinor gauge
\begin{equation}
- \delta_{adS} \theta^{\dot \alpha}= ({\cal K}^{-1})^{\dot\alpha}{}_\alpha
[(\M \coth\M \epsilon)^\alpha +(\theta {\cal K})^\beta \Sigma_0^A
f_{A\beta}^\alpha]\,, \label{adSxSisom1}
\end{equation}
\begin{equation}
- \delta_{adS} x^\mu = \xi^\mu (x) + \left[\frac {\tanh\M/2}{\M}
\epsilon\right]^\beta (\theta {\cal K})^\alpha f^a_{\alpha \beta}  
e_a{}^\mu
(x)
\label{adSxSisom2}
\end{equation}
and the compensating $H$ transformation is
\begin{eqnarray}
\Lambda^{ab}
&=& \ell^{ab} + \left(\frac {\tanh\M/2}{\M} \epsilon\right)^\beta (\theta
{\cal K}) ^\alpha (f^{(ab)}_{\alpha\beta} + e_c^\mu \omega_\mu^{ab}
f^c_{\alpha\beta})\,.
\label{adSxSisom3}
\end{eqnarray}
Here $\epsilon$ is the Killing spinor from (\ref{KS}), ${\cal K}$ is
defined in \eqn{Kmatrix} and can be read off from \eqn{KS}, $\xi^\mu$ is
given in \eqn{killingvectorsadS} and \eqn{killingvectorssphere},
$\ell^{ab}$ in \eqn{compadS} and \eqn{compsphere} and $\Sigma^A_0$  
is given
by \eqn{connection}
and
\begin{eqnarray} (\M^2)^\alpha{}_\beta &=&
f^\alpha_{A\gamma}( \theta {\cal K})^\gamma (\theta {\cal K})^\delta
f_{\delta\beta}^A\,.\label{M}
\end{eqnarray}
Here the formulae have been given in $d$ dimensional language. To get the
explicit form of the structure constants one can then split the
$(d-1,1)$ spinor indices into $(p+1,1)$ and $d-p-2$ spinor indices.  These
are then the indices of the spinor representation of the two bosonic
subalgebras
listed in table~\ref{tab:sol}. Alternatively, as was done in  
\cite{Bernard}
one can rewrite the algebra in $d$-dimensional language. This basis of the
algebra follows naturally from the supergravity solution and this is the
subject of the next section.
\setcounter{equation}{0}
\section{Superisometries from supergravity decomposition of the algebra}
Supergravity superspace describes the maximally supersymmetric vacua in a
way which allows to use the cartesian coordinates in which the metric of
the $adS_{p+2} \times S^{d-p-2}$ space is presented in a form where it is
not a product space
\begin{eqnarray}
ds^2 &=& \left(\frac rR\right)^{2/w}dx^m \eta_{mn} dx^n + \left(\frac
Rr\right)^2 dy^{I} \delta_{IJ} dy^{J}\,.\label{cartmetric}
\end{eqnarray}
This form of the metric is obtained from \eqn{metric} by combining $\{r,
\phi^{m'}\}$ into cartesian $y^I$.  The advantage of these coordinates is
that the R-symmetry $SO(d-p-1)$ acts linearly on them \cite{conf}.  The
Killing vectors are
\begin{eqnarray}
\xi^{m} &=& a^m + \lambda^{mn}_M x_n + \lambda_D x^n + (x^2  
\Lambda_K^m - 2
x^m x\cdot \Lambda_K) + (wR)^2 \left(\frac
Rr\right)^{2/w}\Lambda_K^m\,,\nonumber\\
\xi^{I} &=& - w \Lambda_D(x) y^{I}
+ \Lambda_R^{IJ} y_J\,,
\end{eqnarray}
\par
To obtain the compensating $H$-transformation, we compute the solution to
\begin{equation}
0 = \L_{\hat \xi} e^a - \Lambda^{ab} e_b\,.
\end{equation}
where $e^a$ are the vielbein 1-forms derived from \eqn{metric} and the
indices $a$ split into $\{\u m,\u I\}$.  The parameter of the compensating
$H$-transformation in $(x,y)$ coordinates is given by
\begin{eqnarray}
\ell^{\u m\u n} &=& \Lambda^{mn}_M(x)\, \delta_m^{\u m}\delta_n^{\u
n}\,,\nonumber\\
\ell^{\u m \u I} &=& -2 \frac{(wR)}{r} \left(\frac Rr\right)^{1/w}
\Lambda_K^m y^{I}\, \delta_m^{\u m} \delta_{I}^{\u I}\,,\nonumber\\
\ell^{\u I\u J} &=& \Lambda_R^{IJ}\, \delta_{I}^{\u I}  
\delta_{J}^{\u J}\,.
\end{eqnarray} The Killing-spinors are \begin{eqnarray} \epsilon_-  
&=& (wR)
\left(\frac Rr\right)^{1/2w} \frac {y\cdot \Gamma}r\eta\,,\nonumber\\
\epsilon_+ &=& \left(\frac rR\right)^{1/2w} \left(\epsilon_0 - {x\cdot
\Gamma} \eta\right)\,.
\end{eqnarray}
These equations above give us nice and simple initial conditions but we
have also to solve for the geometry and for the full superisometries.  To
do that we will start with the supergravity superspace.  \par In cartesian
coordinates the on shell superfields, defining the torsions and curvature
are not constant anymore, as in the product space case, where the form
fields are just given by the volume forms of $adS$ space or a sphere, but
are only covariantly constant \cite{KRaj}.  We will therefore start from
supergravity superspace and we will see that apart from the fact  
that we do
not have structure constants as in supercoset case, the rest works  
the same
way as before with structure ``constants", which are only covariantly
constant.  For the 11 dimensional cases the $OSp(8|4)$ and $OSp(6,2|4)$
have been rewritten in terms of the solution to the supergravity  
formfields
F in \cite{Bernard}, which yields the algebra in 11-dimensional language.
Then the supercoset method was used to derived the geometric superfields.
In \cite{C} it was shown that this was completely equivalent with
supergravity superspace.  Here we will develop the exact relation between
the coset superspace and supergravity superspace.
\par
In what follows we show how to write the algebras in terms of supergravity
torsion and curvature components.  The torsion and curvature  
constraints of
the supergravity superspace
\begin{eqnarray}
0 &=& \left(dE^\alpha - \ft14 (\Omega\cdot \gamma E)^\alpha - E^\beta E^a
{\cal T}_{a\beta}^\alpha\right)Q_\alpha\,,\nonumber\\
0 &=& \left(dE^a - E^b \Omega_b{}^a - \ft12E^\beta E^\alpha {\cal
T}_{\alpha\beta}^a \right) P_a\,,\nonumber\\
0 &=& \left(d \Omega^{ab} - \Omega^a{}_c \Omega^{cb} - \ft12 E^d E^c {\cal
R}_{cd}{}^{ab} - \ft12 E^\beta E^\alpha {\cal
R}_{\alpha\beta}{}^{ab}\right) M_{ab}\,,
\label{torsions}
\end{eqnarray}
take the form of Maurer-Cartan equations\eqn{MC}.  We have multiplied the
constraints with generators $\{Q_\alpha, P_a, M_{ab}\}$.  In general this
is not the case, only when the curvatures and torsions are covariantly
constant.  To show this we will first translate to the notation used in
supercoset case
\begin{eqnarray}
L^A &=& \{E^{a}, \Omega^{ab}\}\,,\qquad \mbox{so}\qquad A=\{a,
(ab)\}\nonumber\\
L^\alpha &=& \{ E^{\alpha}\}\,.
\label{dic}
\end{eqnarray}
which come with the generators
\begin{equation}
B_A = \{P_{a}, M_{ab}\}\,,\qquad F_\alpha = \{Q_\alpha\}\,.
\end{equation}
Comparing the supergravity constraints with Maurer Cartan equations
(\ref{MC}) we have the following translation
\begin{eqnarray}
\label{transl1}
f^a_{\alpha\beta} &\rightarrow& - {\cal T}^a_{\alpha\beta}\,,\\
f^{(ab)}_{\alpha\beta} &\rightarrow& - \R^{ab}_{\alpha\beta}\,,\\
f^{(ab)}_{cd} &\rightarrow& - R_{cd}{}^{ab}\,,\\ f^\alpha_{a\beta}
&\rightarrow& - \T^\alpha_{a\beta}\,,\\
f^{\alpha}_{(ab)\beta} &\rightarrow& \ft14 (\Gamma_{ab})^{\alpha}{}_\beta
\\ f^{a}_{(cd) b} &\rightarrow& - \eta_{cb} \delta^a_b \,,\\
f^{(ab)}_{(el) \; (cd)} &\rightarrow& \delta^a_{e} \eta_{lc}  
\delta^b_d \,.
\label{trans}
\end{eqnarray}
Thus we may rewrite (\ref{torsions}) as Maurer Cartan equations with soft
structure ``constants" $f^\Lambda_{\Sigma \Delta}(\Phi^\alpha)$ depending
on some covariant superfields $\Phi^\alpha$\footnote{These superfields are
the forms, the dilaton and dilatino of the background.  As the  
dilaton is a
scalar this requires it to be constant, which occurs for the $D3$-brane
solution and therefore for $D3$ the superspace is a coset superspace.},
whose properties are defined in the Appendix A.2.
\begin{eqnarray}
\left(d L^\Lambda + \frac 12 L^\Delta \wedge L^\Sigma
f_{\Sigma\Delta}{}^\Lambda (\Phi) \right) {\bf T}_\Lambda =0\,.
\end{eqnarray}
This is the statement that there exists a nilpotent differential operator
\begin{equation}
{\cal D} = d - L^\Lambda T_{\Lambda} \qquad {\cal D}^2=0
\end{equation}
and the generators of the soft algebra satisfy $[\, {\bf T}_\Lambda , {\bf
T}_\Sigma\, \}= f_{\Lambda\Sigma}{}^\Delta (\Phi^\alpha) {\bf T}_\Delta$.
This is only possible iff
\begin{equation}
{\cal D} f_{\Sigma\Delta}{}^\Lambda (\Phi^\alpha) \equiv d
f_{\Sigma\Delta}{}^\Lambda - \Phi^\beta (L^\Pi {\bf T}_\Pi)_\beta{}^\alpha
\partial_\alpha f_{\Sigma\Delta}^{\Lambda} =0
\end{equation}
since
${\cal D}^2 f_{\Sigma\Delta}{}^\Lambda (\Phi^\alpha)=0$.  This is in fact
equivalent to the statement that the structure ``constants" of the soft
algebra have to satisfy the following generalized Jacobi identities:
\begin{equation}
0=\Sigma_3^\Sigma \Sigma_2^\Delta(f_{\Delta\Sigma}{}^\Pi
\Sigma_1^\Gamma f_{\Gamma\Pi}{}^\Lambda - \Phi^\beta (\Sigma_1^\Pi {\bf
T}_\Pi)_\beta{}^\alpha \partial_\alpha f_{\Delta\Sigma}{}^\Lambda) +
\mbox{cyclic}(1\rightarrow2\rightarrow3)\,.
\label{Jacobis}
\end{equation}
Now we may use the dictionary (\ref{transl1})-(\ref{trans}) to  
identify the
soft algebra with structure functions depending on curvatures and torsions
of supergravity.  There is the universal part containing $M$, which  
will be
$H$
\begin{eqnarray}
{}[M_{ab}, M_{cd}] &=& \eta_{a[c} M_{d]b} - \eta_{b[c}
M_{d]b}\,,\nonumber\\
{}[P_a, M_{bc}] &=& \eta_{a[b} P_{c]}\,,\qquad
[M_{ab}, Q_\alpha] = -\frac14 (\Gamma_{ab} Q )_\alpha\, \label{STABM}
\end{eqnarray}
and
\begin{eqnarray}
{}[P_a, P_b] &=& - {\cal R}_{ab}{}^{cd} M_{cd}\,,\label{PPcom}\\
{}[P_a, Q_\alpha] &=& - {\cal T}_{a\alpha}{}^\beta Q_\beta\,,\nonumber\\
{}\{Q_\alpha, Q_\beta\} &=& - {\cal T}_{\alpha\beta}{}^a P_a - {\cal
R}_{\alpha\beta}{}^{ab} M_{ab}\,.
\end{eqnarray}
One should read these formulas carefully.  In the case that one has a
product geometry as is $adS \times S$ the curvature ${\cal R}_{ab}{}^{cd}$
will split into two parts and therefore the generators $M_{ab}$ which are
off-diagonal, i.e.  they have an index in the tangent space of each of the
two factors in the product space, will not appear on the r.h.s.  of
\eqn{PPcom}.  For consistency of the algebra therefore these generators
should also be dropped in \eqn{STABM}.  An example of this is that the
tangent space group of $adS_{p+2}\times S^{d-p-2}$ is $SO(p+1,1)\times
SO(d-p-2)$ in stead of the full $SO(d,1)$ as in flat space.  \par For flat
space we have
\begin{equation}
{\cal T}_{\alpha\beta}^a = (\Gamma^a)_{\alpha\beta}\,, \qquad{\cal
T}_{\alpha a}^\beta = {\cal R} = 0 \,.
\end{equation}
For the general case the solution is given by
\begin{eqnarray}
E^\alpha &=& \left(\frac{\sinh \M}{\M}\right)^{\alpha}_{\beta} (D
\Theta)^\beta\,,\\ E^a &=& e^a - \Theta^\alpha \T_{\alpha\beta}^a
\left(\frac{\sinh\M/2}{\M^2}\right)^\beta_\gamma (D\Theta)^\gamma\,,\\
\Omega^{ab} &=& \omega^{ab} - \Theta^\alpha \R_{\alpha\beta}^{ab}
\left(\frac{\sinh\M/2}{\M^2}\right)^\beta_\gamma (D\Theta)^\gamma\,,
\end{eqnarray}
\begin{equation}
({\cal M}^2)^\alpha_\beta = - \ft14 (\gamma_{ab})^\alpha{}_\gamma
\Theta^\gamma \Theta^\delta \R_{\delta\beta}^{(ab)} -\T^{\alpha}_{a  
\gamma}
\Theta^\gamma \Theta^\delta \T^a_{\delta\beta}\,.
\end{equation}
One can verify that these results in terms of the structure functions,
using the dictionary (\ref{transl1})-(\ref{trans}) are the same as the one
in the supercoset space We find the superisometries the same way as before
and therefore the result is given \eqn{adSxSisom1}-\eqn{adSxSisom3} where
the dictionary (\ref{transl1})-(\ref{trans}) has to be used to replace the
expressions for constant $f^\Lambda_{\Sigma \Delta}$ by their supergravity
counterparts which are covariantly constant.
\setcounter{equation}{0}
\section{Discussion}
Thus we have found the superisometries, the combination of the
transformations of coordinates $(X, \theta)$ of the near-horizon  
superspace
and the compensating Lorentz transformations, under which the  
vielbeins and
the spin connection of the superspace are invariant.  In the picture  
of the
product space $adS_{p+2}\times S^{d-p-2}$ where a ${G\over H}$ supercoset
construction is available \cite{KRR}, the isometries are given in
\eqn{adSxSisom1}-\eqn{adSxSisom3}.  The result covers all cases of the
superconformal algebras ${\bf G}$.  The data required to construct the
whole superspace and its isometries consists of the the algebra and
the Killing spinors and vectors at $\theta=0$, which
have been discussed in sec.~4.2. Alternatively one can start from
supergravity, i.e.~one takes a solution to the supergravity field
equations (with vanishing fermions) and the Killing vectors
and Killing spinors and compensating Lorentz transformation can be  
obtained
as the transformations that leave the solution invariant.
To construct the higher order $\theta$ components one identifies
the supertorsion and curvature components with the structure  
`constants' of
the algebra, the dictionary has been given in \eqn{transl1}-\eqn{trans},
and we can apply our general formula in terms of structure constants.  It
has been observed that for the maximally supersymmetric solutions with
vanishing fermions the supertorsions and curvatures are only covariantly
constant \cite{KRaj}. This happens for instance if the near-horizon
geometry is given in cartesian coordinates.  Fortunately our construction
of the superisometries thus not demand structure constants, but only
covariantly constant structure functions.  The derivation of this result
can be found in sec.~5.
\par
Having established the full set of isometries in $adS$ superspace, we may
use them for few applications, in particular we would like to study the
condition of the BPS states on the brane in an $adS$ background.
\leftmargini=5mm
\begin{enumerate}
\item {\sl BPS States of Branes in adS.}\\
The classical worldvolume action has two types of fermionic symmetries: a
global one, the fermionic isometry of the flat superspace and a local one,
the so-called $\kappa$-symmetry.  On the fermionic variables $\theta$ they
act as
\begin{equation}
\delta_f \theta = -\epsilon + (1+\Gamma) \kappa\,.
\end{equation}
Here $\Gamma$ defines the $\kappa$-symmetry depends in general on
$(X,\theta)$ and on the worldvolume forms $F$, i.e.~we have $\Gamma (X,
\Theta, F) $.  The matrix $\Gamma$ satisfies $\Gamma^2=1$ and $\mbox{tr}\,
\Gamma=0$.  To identify the equation which defines some  
configuration to be
a BPS configuration of the brane in flat space, one has to require  
that the
total transformation of the fermionic variables vanishes at vanishing
fermions
\begin{equation}
\delta_{BPS} \theta \equiv (\delta \theta)_{\theta=0} = \left ( - \epsilon
+ (1+\Gamma ) \kappa \right)_{\theta=0} = 0\,.
\label{BPS}
\end{equation}
Now we may observe that for vanishing fermions
\begin{equation}
\delta_{BPS} (1 - \Gamma) \theta = - (1-\Gamma)_{\theta=0} \epsilon = 0
\end{equation} due to the fact that $(1-\Gamma )(1+\Gamma )=0$.
This form for the preservation of supersymmetry was first discussed
in~\cite{bbs} and was subsequently considered in \cite{oog,bkop,glw,gp}.
The number of independent zero modes $\epsilon$ of this equation can  
be 1/2
or less than the total number of components of fermions $\theta$.
\par
What from this analysis carries over to the case when the branes are in
$adS$ superspace?  One has to start with the $\kappa$-symmetric brane
actions in the $adS$-superspace background and formulate the total
supersymmetry transformation on the fermions.  It consists of the  
fermionic
superisometry of the background and of a local $\kappa$-symmetry.
\begin{equation}
(\delta\theta^{\dot \alpha}) E_{\dot \alpha} {}^\alpha (X, \theta)=
(\delta_{adS} \theta^{\dot \alpha}) E_{\dot \alpha} {}^\alpha (X,  
\theta) +
(1+\Gamma)\kappa\,.
\label{total}
\end{equation}
Here we have taken into account that in the Killing spinor gauge the
gravitino superfield vanishes, $E_\mu^\alpha(X, \Theta)=0$.  The condition
of unbroken supersymmetry of the BPS state on the brane at $\theta=\delta
\theta=0$ is reduced to the condition
\begin{equation}
\Sigma^\alpha (X, \theta=0) +(1+\Gamma(X, \theta=0, F) )\kappa =0\,.
\end{equation}
One can as before multiply this condition on $(1-\Gamma(X, \theta=0, F) )$
and get\footnote{This condition was also suggested in \cite{BC}}
\begin{equation}
\left (1-\Gamma(X, F)\right) ^\alpha _\beta \Sigma_0^\beta
(X) =0\,.
\label{unbrBPS}
\end{equation}
Here $\Sigma_0^\beta (X)$ as it follows from our complete set of
transformations, is the Killing spinor of the background.  Note also that
$\kappa$-symmetry transformation carries the information about the brane
action whose symmetries are investigated as well as the information on the
background.  In particular one can look at any D-p-brane of IIB theory in
$adS_5\times S^5$ background and find all possible BPS states.  For  
example
one can study a D5 brane in the near horizon background of D3 brane.  The
issue of BPS states of such configuration was studied in \cite{Cal}
following \cite{Imamura}.  The derivation of the BPS condition there is
surprisingly complicated and never actually using the direct supersymmetry
of the D5 brane in $adS_5\times S^5$ background which is a combination of
the background isometries and $\kappa$-symmetry.  Thus we would like to
stress here that (\ref{unbrBPS}) may have many possible solutions,
only small part of which was recently studied in \cite{BC}.  \item{\sl Use
of superisometries for GS Superstring in adS background}\\ A classical GS
Superstring in the near horizon superspace of the D3 branes has the
combinations of symmetries as shown in (\ref{total}).
\begin{equation}
(\delta\theta^{\dot \alpha}) E_{\dot \alpha} {}^\alpha (X, \theta)=
(\delta_{adS^5\times S^5} \theta^{\dot \alpha}) E_{\dot \alpha} {}^\alpha
(X, \theta) + (1+\Gamma_{string})\kappa
\label{totalGS}
\end{equation}
supplemented with the corresponding transformations of the bosonic fields.
Here the superspace vielbeins $E_{\dot \alpha} {}^\alpha (X, \theta)$ are
defined for the near horizon D3 geometry, and the generator of a local
$\kappa$-symmetry is the one for the string in the D3 background.  The
expression for the matrix ${\cal M}^2$ simplifies in any of the gauges
$\Theta_+=0$ or $\Theta_-=0$, it becomes a nilpotent matrix and therefore
squares to zero.  An easy way to see this is to observe that
\begin{equation}
{\cal M}^4 A = [\Theta [\Theta [\Theta [\Theta , A]]]]_{\Theta_{\pm}=0} =0
\end{equation}
and therefore
\begin{equation}
({\cal M}^4 A )_{\Theta_{\pm}=0} =0
\end{equation} For $\Theta_+=0$
or $\Theta_-=0$ the 4 objects with $\Theta$ accumulate the scaling weight
$4\times \pm {1\over 2}$ and therefore together with the scaling weight of
the fermionic operator in $A$ the weight of the operator in this multiple
commutator becomes equal to $\pm 2 \pm {1\over 2}$, i.e.  $\pm  
{5\over 2}$.
Such operators do not exist in our algebras and therefore this expression
vanishes for arbitrary $A$.  Note that in case $\Theta_-=0$ a stronger
restriction is available.  In addition to $\Theta_-=0$ one can show that
also $(D\Theta)_-=0$ and in such case even ${\cal M}^2 D\Theta=0$.  Indeed
here we use
\begin{equation}
{\cal M}^2 D\Theta = [\Theta [\Theta , D\Theta]
\end{equation}
and therefore
\begin{equation}
({\cal M}^2 D\Theta)_{\Theta_{-}=0} =0
\end{equation}
as the weight of all 3 fermionic operators sums up to $3/2$.  Here  
again we
take into account that the fermion operators with such weight are not
available in our algebras.  This leads to the following simplification of
the isometries in the gauge-fixed form of the action.  In  
particular, if we
choose, as in \cite{GS}, the gauge $\Theta_- = \theta_-=0$, we find that
the contribution from isometries to the total transformations is
simplified:
\begin{equation}
-\delta \theta_- = \epsilon_{0-} +\Delta(\kappa ) =0
\end{equation}
and
\begin{equation}
-\delta \theta_+ = \epsilon_{0+} +{1\over 2} ({\cal
K}^{-1} {\cal M}^2 {\cal K})_{+-} \epsilon_{0- } + \Delta(\kappa)
\end{equation}
where $\epsilon_{\pm0}$ are constant spinors and $\Delta(\kappa)$ is the
contribution from the local $\kappa$-symmetry.  The second term in
$\delta\theta_+$ is only quadratic in fermionic variables.  These  
equations
give the basis for the derivation of the symmetries of the gauge-fixed IIB
string action in adS background \cite{GS}.  From these symmetries one
should be able to derive the Ward Identities which will be responsible for
the properties of the loop corrections.  Since the rigid symmetries of the
gauge-fixed action will form the superconformal algebra, the spectrum of
states must also form a representations of this algebra.  This  
symmetry may
help us to find the spectrum of states of the string in adS.
\item {\sl Boundary limit}\\
One may try to use the superconformal symmetry
of the near horizon superspace to find the limit to the boundary of the
$adS$ space, as suggested to us by S. Shenker.  This will provide a  
particular
realization of the superconformal
algebras closely related to the properties of the conformal field theories
at the boundary.  It may also help to clarify the $adS$/CFT correspondence
conjectured by Maldacena. This project is currently under investigation in
\cite{CKR}. One of the striking new results in this study is the  
derivation of
the off-shell harmonic superspace of super Yang-Mills theory from  
the boundary
limit of the supergravity/IIB string compactified on $adS_5\times S^5$.
\end{enumerate}
\medskip
\section*{Acknowledgments.}
\noindent
We had useful discussions with I.~Antoniadis, J.~Bagger, J.~Gomis,
J.~Maldacena,
J.~Rahmfeld, S.~Shenker, L.~Susskind, A.~Tseytlin  and Y.~Zunger.   
The work of
R.~K.  is supported by the NSF
grant PHY-9870115.  The work of P.~C.  is supported by the European
Commission TMR programme ERBFMRX-CT96-0045.  We are grateful to the
organizers of the mid-term TMR meeting and the Summer School in  
Corfu where
this work was initiated.
\newpage
\appendix
\setcounter{equation}{0}
\section{Conventions}
\subsection{Super differential forms}
To define differential forms we use the basic rule
\begin{equation}
dZ^M \wedge dZ^N = - (-)^{MN} dZ^{N} \wedge dZ^M\,,
\end{equation}
where $(-)^{MN}=1$ unless both indices are spinorial.  The  
components of an
$n$-form $\Phi_n$ are defined through
\begin{equation}
\Phi_n= \frac1{n!} dZ^{M_1}\wedge\dots\wedge dZ^{M_p} \Phi_{M_p\dots
M_1}\,.
\end{equation}
The exterior derivative $d$ is given by
\begin{equation} d\Phi_n = \frac1{n!}dZ^{M_1} \wedge \dots\wedge
dZ^{M_p}\wedge dZ^{N} \partial_N \phi_{M_p\dots M_1}\,,
\end{equation}
and is nilpotent
\begin{equation}
d^2 = 0\,.
\end{equation}
$d$ starts from the right which means that
\begin{equation}
d(\Phi_m\wedge \Phi_n) = \Phi_m\wedge (d\Phi_n) + (-)^n
(d\Phi_m)\wedge \Phi_n\,.
\end{equation}
Also we take the convention that $d$ commutes with $\theta^A$.
\par
A useful second operator is the Lie derivative along a vector field
$\Xi=\Xi^M \frac{\partial}{\partial Z^M}$ of an $n$-form $\Phi_n$
\begin{equation}
{\cal L}_\Xi \Phi_n = \frac1{n!} dZ^{M_1}\wedge\dots\wedge
dZ^{M_n} \left(\Xi^{N} \partial_N \Phi_{M_n\dots M_1} + n (\partial_{M_n}
\Xi^{N}) \Phi_{NM_{(n-1)}\dots M_1}\right)\,.
\end{equation}
\subsection{Gauging (soft) superalgebras}
We consider a supergroup $G$ with associated superalgebra {\bf
G}.\footnote{This section is based on \cite{Toinekar} but written in form
notation.} Consider a bunch of `covariant' fields $\Phi^\alpha$, they
transform in some representation labeled by $\alpha$ of $G$.  The
generators of the algebra are ${\bf T}_\Lambda$ and satisfy
\begin{equation}{}
[\, {\bf T}_\Lambda , {\bf T}_\Sigma\, \} \equiv {\bf T}_\Lambda {\bf
T}_\Sigma \mp {\bf T}_\Sigma {\bf T}_\Lambda = f_{\Lambda\Sigma}{}^\Delta
(\Phi^\alpha) {\bf T}_\Delta\,,
\label{app:Galg}
\end{equation}
where the $+$ sign is taken if both $\Lambda$ and $\Sigma$ are fermionic
indices.  We have included the possibility of {\sl soft} algebras, where
the structure `constants' are actually structure functions.
\par
We define a {\bf G} valued object $A$ as
\begin{equation}
A=A^\Lambda {\bf T}_\Sigma\,.
\end{equation}
It follows that for two bosonic {\bf G} valued objects $A$ and $B$
\begin{equation}
[\, A,B\,] = [\, A^\Lambda {\bf T}_\Lambda, B^\Sigma {\bf
T}_\Sigma\,] = B^\Sigma A^\Lambda [\, {\bf T}_\Lambda,{\bf T}_\Sigma\,\} =
B^\Sigma A^\Lambda f_{\Lambda\Sigma}{}^\Delta {\bf T}_\Delta\,.
\end{equation}
\par
We will consider the ${\bf T}$ as ``active'' operators.  Take, e.g.~on the
fields field $\Phi^\alpha$ an infinitesimal group transformation is  
denoted
by
\begin{equation}
\delta(\Sigma) \Phi^\alpha = \Phi^\beta \Sigma^\Lambda
{\bf T_\Lambda}_\beta{}^\alpha \equiv \Phi^\beta
\Sigma_\alpha{}^\beta\,,\quad \mbox{where}\quad \Sigma = \Sigma^\Lambda
{\bf T}_\Lambda\,.
\end{equation}
The infinitesimal parameters $\Sigma$ are {\bf G} valued objects.  If we
apply a second variation it acts as
\begin{equation}
\delta(\Sigma_1) \delta(\Sigma_2) \Phi^\alpha = \Phi^\gamma
\Sigma_2{}_\gamma{}^\beta \Sigma_1{}_\beta{}^\alpha\,,
\end{equation}
thus $T_{\Lambda_1}$ works on $(\Phi T_{\Lambda_2})$.
\par
Now we can look at the commutator of two such transformations on a
covariant field.  It follows from \eqn{app:Galg} that they should satisfy
\begin{equation}
{}[\delta(\Sigma_1), \delta(\Sigma_2)] = \delta (\Sigma_2^\Delta
\Sigma_1^\Pi f_{\Pi\Delta}{}^\Lambda)\,.
\label{transalg}
\end{equation}
We have to take into account that the structure constants also transform
\begin{equation}
\delta (\Sigma) f_{\Sigma\Delta}{}^\Lambda = \Phi^\beta
\Sigma_{\beta}{}^\alpha \frac\partial{\partial\Phi^\alpha}
f_{\Sigma\Delta}{}^\Lambda\,.
\end{equation}
The Jacobi identities which follow from $0 =
[\delta(\Sigma_1),[\delta(\Sigma_2),\delta(\Sigma_3)] +
\mbox{cyclic}(1\rightarrow2\rightarrow3)$ are then
\begin{equation}
0=\Sigma_3^\Sigma \Sigma_2^\Delta(f_{\Delta\Sigma}{}^\Pi \Sigma_1^\Gamma
f_{\Gamma\Pi}{}^\Lambda - \delta(\Sigma_1) f_{\Delta\Sigma}{}^\Lambda) +
\mbox{cyclic}(1\rightarrow2\rightarrow3)\,.  \label{app:Jacobis}
\end{equation}
We want to gauge the superalgebra {\bf G}.  The parameters $\Sigma$  
are now
coordinate dependent $\Sigma(Z)$.  The gauge field $A$ is a {\bf G} valued
1-form.  We introduce a covariant derivative denoted by
\begin{equation}
{\cal D} = d - A\,.
\end{equation}
On a $G$-covariant field it acts as
\begin{equation}
{\cal D} \Phi^\alpha = d\Phi^\alpha - \Phi^\beta A_\beta{}^\alpha\,.
\label{Doncov}
\end{equation}
It acts {\sl from the right}, consistent with the convention that $d$ acts
from the right.  We define the transformation of the gauge field $A$, such
that in $\delta(\Sigma) ({\cal D}\Phi)^\alpha$ there is no derivative on
the parameters $\Sigma^\Lambda$.  This can be achieved by taking
\begin{eqnarray}
\delta(\Sigma) A &=& d\Sigma + [\, A, \Sigma\,]\nonumber\\
&=& \left( d\Sigma^\Lambda + \Sigma^\Delta A^\Sigma
f_{\Sigma\Delta}{}^\Lambda\right){\bf T}_\Lambda\,.
\label{app:gauge}
\end{eqnarray}
Trying to close the commutator \eqn{transalg} on $A$, we find that
\begin{equation}
{\cal D} f_{\Sigma\Delta}{}^{\Lambda} = d f_{\Sigma\Delta}{}^\Lambda -
\Phi^\beta (A^\Pi {\bf T}_\Pi)_\beta{}^\alpha \partial_\alpha
f_{\Sigma\Delta}^{\Lambda}
\label{covconst}
\end{equation}
Therefore one can gauge the {\sl soft} algebras if the structure functions
are covariantly constant.
\par
One can construct the $G$-covariant curvature $F$, which is a {\bf G}
valued 2 form, by
\begin{equation}
{\cal D}^2 = -F\,
\end{equation}
and it follows that
\begin{eqnarray}
F &=& d A - A\wedge A\nonumber\\
&=& \left(d A^\Lambda + \frac 12 A^\Delta \wedge A^\Sigma
f_{\Sigma\Delta}{}^\Lambda\right) {\bf T}_\Lambda\,.
\end{eqnarray}
$F$ transforms in the adjoint (only if \eqn{covconst} is satisfied), i.e.
\begin{equation}
\delta(\Sigma) F = [F,\Sigma] = - F^\Sigma \Sigma^\Delta
f_{\Delta\Sigma}{}^\Lambda {\bf T}_\Lambda\,.
\end{equation}
The adjoint representation matrices are given by
\begin{equation}
({\bf T_\Lambda})_\Sigma{}^\Delta = -f_{\Lambda\Sigma}{}^\Delta
\end{equation}
and satisfy \eqn{Galg} by virtue of the Jacobi identities.
The curvatures satisfy the Bianchi identities (again only if  
\eqn{covconst}
is satisfied)
\begin{equation}
0 = {\cal D} F = dF + [A,F]\,,
\end{equation}
using \eqn{Doncov}.
\setcounter{equation}{0}
\section{Coset manifolds and superisometries \label{app:coset}}
Consider arbitrary elements $g$ and $h$ of the groups $G$ and $H$
respectively.  We define equivalence classes in $G$: two elements $g$ and
$g'$ belong to the same equivalence class iff they can be connected by a
right multiplication with an element of $H$, i.e.
\begin{equation}
g = g'\, h\,.
\end{equation}
This equivalence class is called the left coset of $g$.  The set of all
cosets is the coset manifold denoted by $G/H$.
\par
Now we can characterize each coset by a coset representative ${\rep}(Z)$,
labelled by as many coordinates $Z$ as we need, typically for the
supercoset spaces we consider $Z= \{ X^\mu,\theta^{\dot \alpha} \}$.  It
parametrizes the coset manifold if each coset contains exactly one of the
$\rep(Z)$'s.  Once we have chosen a representative it is clear that every
group element $g$ can be decomposed into
\begin{equation}
g = \rep(Z) h\,,
\end{equation}
where $\rep(Z)$ is the representative of the coset to which
$g$ belongs and $h$ acts in this coset.
\par
Now it is clear that a product with an arbitrary group element $g$ of $G$
with a coset representative $\rep(Z)$ can bring you to another coset and
\begin{equation}
g \rep(Z) = \rep(Z') h\,,
\label{grouptrans}
\end{equation}
where in general
\begin{equation}
Z' = Z'(g,Z)\,,\qquad h=h(g,Z)\,.
\end{equation}
The {\sl global} group transformations $g$ correspond to the
(super)isometry group of the (super)coset space.
\par
We define the left-invariant Cartan 1-forms
\begin{equation}
{\rep}(Z)^{-1} d {\rep}(Z) = L(Z)\,.
\end{equation}
Since $L(Z)$ is a group element close to the identity it is a {\bf G}
valued super 1-form
\begin{equation}
L = L^\Lambda {\bf T}_\Lambda = dZ^M L_M{}^\Lambda {\bf T}_\Lambda\,.
\end{equation}
They are invariant under a {\sl global} $G$-transformation
from the left
\begin{equation}
{\rep}(Z) \rightarrow g{\rep}(Z)\,.
\end{equation}
Indeed, we have
\begin{equation}
(g {\rep}(Z))^{-1} d (g {\rep}(Z)) - {\rep}(Z)^{-1} d {\rep}(Z) =
{\rep}(Z)^{-1} g^{-1} g d{\rep}(Z) - {\rep}(Z)^{-1} d {\rep}(Z) = 0\,.
\label{leftinv}
\end{equation}
\par
The Cartan 1-forms satisfy the Maurer Cartan equations (\ref{MC}) which is
easily deduced by substituting $L = \rep^{-1} d \rep$ into this equation.
\par
We can rewrite \eqn{leftinv} by using \eqn{grouptrans} and obtain
\begin{equation}
0 = h^{-1} L(Z') h + h^{-1}dh - L(Z)\,.\label{kill1}
\end{equation}
Now we are interested in infinitesimal transformations of
the coordinates
\begin{equation}
Z'{}^{ M} = Z^{ M} - \Xi^{ M}(Z)\,.
\end{equation}
The compensating $H$-transformation are infinitesimal and we
define
\begin{equation}
h = 1 - \Lambda = 1 - \Lambda^i {\bf H}_i\,.
\end{equation}
One easily derives that
\begin{equation}
L(Z') = L(Z) - \L_{\Xi} L(Z)\,.
\end{equation}
Therefore \eqn{kill1} becomes
\begin{eqnarray} 0 &=& \L_\Xi L + d\Lambda + [\, L(Z),
\Lambda\,]\nonumber\\
&=& \left(\L_\Xi L^\Lambda + d\Lambda^i \delta_i{}^\Lambda + \Lambda^i
L^\Sigma f_{\Sigma i}{}^\Lambda\right) {\bf T}_\Lambda\,.
\end{eqnarray}

\end{document}